\documentclass{nature}

\usepackage{graphicx}
\makeatletter
\let\saved@includegraphics\includegraphics
\AtBeginDocument{\let\includegraphics\saved@includegraphics}
\renewenvironment*{figure}{\@float{figure}}{\end@float}
\makeatother

\usepackage{soul}
\usepackage{amsmath}
\usepackage{siunitx}
\usepackage{tocloft}
\usepackage{tikz}
\usepackage{pdfpages}
\usepackage{upgreek}
\usepackage{enumitem}
\usepackage{float}
\usepackage{amssymb}
\usepackage{makeidx}
\usepackage{multirow}
\usepackage{indentfirst}
\usepackage[toc,page]{appendix}
\usepackage{lipsum}
\usepackage[multiple]{footmisc}
\usepackage[utf8]{inputenc}
\usepackage{mathtools}
\usepackage{amsbsy,bm}

 \newcommand{\be}{\begin{equation}}
 \newcommand{\ee}{\end{equation}}
 \newcommand{\bea}{\begin{eqnarray}}
 \newcommand{\eea}{\end{eqnarray}}

\title{Unsteady wetting of soft solids}

\author{Surjyasish Mitra$^{1\,*}$, Quoc Vo$^{2\,*}$ , Marcus Lin$^2$ \& Tuan Tran$^2$}

\begin{document}

\maketitle

\begin{affiliations}
 \item School of Physical and Mathematical Sciences, 
 Nanyang Technological University, 
 50 Nanyang Avenue, 639798, Singapore.
 \item School of Mechanical \& Aerospace Engineering, 
	Nanyang Technological University, 
	50 Nanyang Avenue, 639798, Singapore.\\
$^*$These authors contributed equally.	
\end{affiliations}

\begin{abstract}
From hydrogels and plastics to liquid crystals, 
soft solids cover a wide array of synthetic and biological
materials \cite{jones2002soft} that play key enabling roles 
in advanced technologies such as 
3D printing \cite{truby2016printing},
soft robotics \cite{park2016phototactic}, 
wearable electronics \cite{ma2013bio}, 
self-assembly \cite{kato2006functional},
and bioartificial tissues \cite{zhao2011active,fusco2014integrated}.
Their elasticity and stimuli-induced changes
in mechanical, optical, or electrical properties
offer an unique advantage
in designing and creating new dynamically functional 
components for sensing, micro-actuation, colour changes, information 
and mass transport. 
To harness the vast potential of soft solids 
through their ability to respond to the environment,
a thorough
understanding of their 
reactions when 
exposed to liquids is needed. 
Attempts to study the interactions between 
soft solids
and liquids have largely focused on 
the wetting of soft solids
\cite{lester1961contact,lester1966contact,style2017elastocapillarity,chen2017static,andreotti2020statics} 
and its resulting 
deformation
at equilibrium \cite{style2013universal,andreotti2020statics},
in quasi-static state \cite{gerber2019wetting}, 
or in steady state \cite{kajiya2013advancing,kajiya2014liquid}.
Here we consider the frequently encountered case
of unsteady wetting of a liquid on a soft solid,
and show that 
transient deformation of the solid
is necessary to understand 
unsteady wetting behaviours.
We find that the initial spreading of the liquid
occurs uninterupted
in the absence of solid deformation.
This is followed by intermittent spreading,
in which transient deformation of 
the solid at the three-phase contact line (CL) causes 
alternation between CL 
sticking and slipping.
We identify the spreading rate of liquids and the 
viscoelastic reacting rate of soft solids 
the two competing factors in 
initiating intermittent
spreading. 
We formulate and validate experimentally
the conditions required
for the contact line to transition from sticking 
to slipping. 
By considering the growing deformation of soft solids 
as dynamic surface heterogeneities,
our proposed conditions for stick-slip transition in 
unsteady wetting on soft solids
broaden the classical theory on 
wetting hysteresis on rigid solids.
Our results provide a basis to understand
dynamic responses of soft solids to  
unsteady wetting.
\end{abstract}

Wetting phenomena on soft solids
has been studied extensively for the past few decades
\cite{lester1961contact,lester1966contact,style2017elastocapillarity,chen2017static,andreotti2020statics}
owning to the
increasingly central role of soft materials
in advanced applications, from  
flexible electronics \cite{ma2013bio} and 
soft robotics \cite{park2016phototactic} 
to biomedicine \cite{zhao2011active,fusco2014integrated}. 
Majority of research focuses on 
wetting behaviours with the liquid spreading rate 
smaller than the 
solid reacting rate against  
the liquid-air interfacial 
tension at the spreading front.  
These behaviours are often observed 
at equilibrium \cite{style2013universal,andreotti2020statics}, 
in quasi-static state \cite{gerber2019wetting},
or in steady state by forced contact line (CL) motion \cite{kajiya2013advancing,kajiya2014liquid}.
In contrast, unsteady wetting on soft solids
is a more frequently encountered 
but much less explored situation, 
where unsteady liquid spreading is spontaneously driven 
by 
surface
tension imbalance at the contact line
without any external control on 
the spreading rate. 
In this case, CL displacement and 
formation of the so-called 
wetting ridge, i.e., solid deformation at the contact line,
are both in transient states.
The resulting intermittent CL motion  
in unsteady wetting is largely unexplored, 
as opposed to its counter-part in quasi-static or forced wetting
\cite{kajiya2013advancing,kajiya2014liquid,van2018dynamic,gerber2019wetting,park2017self,guan2020state}, 
and is therefore the focus of our study.

In this Article, we investigate 
intermittent spreading behaviour
of unsteady wetting on soft solids. 
We let liquid droplets to spread freely
when they are deposited on soft surfaces, 
thereby allowing the liquid to spread and interact with 
the developing wetting ridge without any external constraint. 
We find that for a wide range of liquid's surface tension and 
solid's elasticity,
fast inertial spreading transitions to 
intermittent spreading, i.e., alternation between CL sticking and slipping,
when the spreading rate reduces and 
becomes
comparable to the
viscoelastic reacting rate of the soft solid. 
The stick-slip spreading behaviour in this case 
consists of two essential elements, liquid spreading and 
wetting ridge formation, both in their transient states. 
Whenever the contact line (CL) sticks, 
it liberates itself and slips 
due to the combined effect of capillary force and flow inertia accumulated 
from the early inertial spreading.

\section*{Results}

\begin{figure}
\includegraphics[width=1\textwidth]{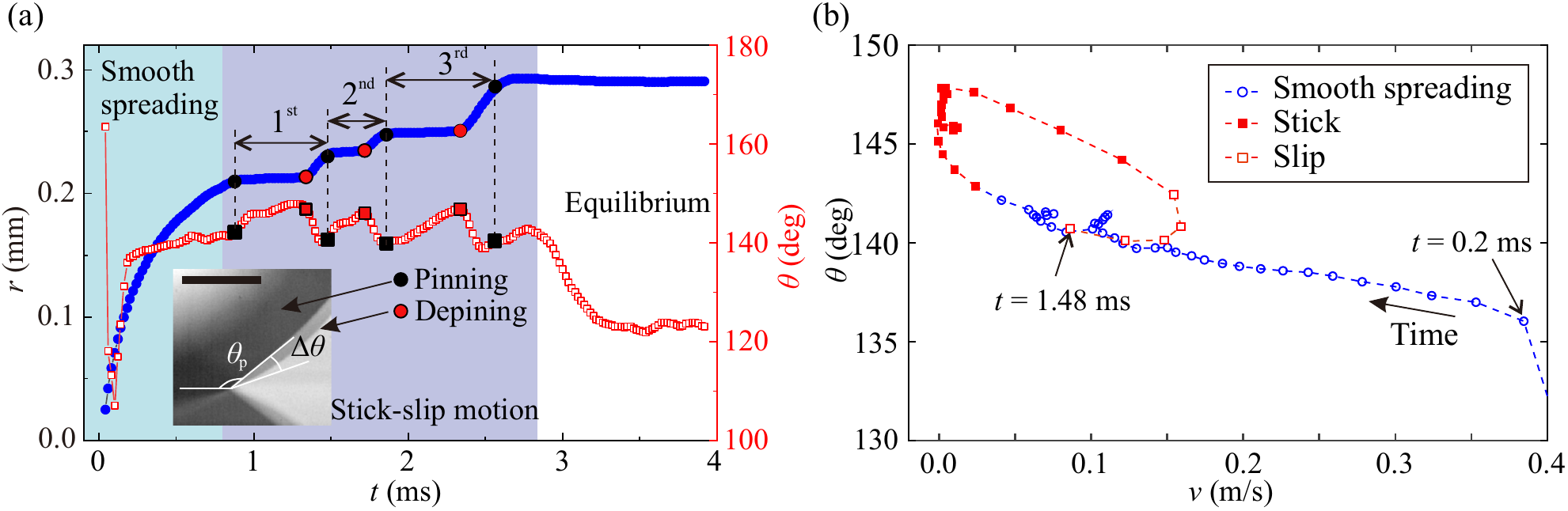}
\centering
\caption{\small{
(a) Variation of droplet - solid footprint radius, i.e., the spreading radius 
$r$ and dynamic contact angle 
$\theta$ 
over time $t$ 
for a 0.5\,mm 
radius water droplet spreading on
a soft surface with $G = 1.6\,{\rm kPa}$.
Inset: a superimposed image of 
the pinning 
and depinning 
snapshots
of the third 
stick-slip event
in the main plot.
The scale bar represents 
100$\,\upmu$m.
(b) Plot showing the dependence of dynamic contact angle $\theta$
on the velocity $v$ 
of the contact line for experiment shown in (a).
Only the data corresponding to the smooth spreading 
phase (blue open circles), the first stick phase (closed red squares)
and the first slip phase (open red squares) are shown.
}}
\label{fig:1}
\end{figure}

We perform numerous 
unsteady wetting experiments on soft substrates
and characterise 
the spreading dynamics of liquids, focusing on 
the stick-slip behaviour before the liquids reach
equilibrium (Methods).  
We 
track
the contact radius $r$
and the corresponding 
contact angle
$\theta$ 
with respect to time $t$
for liquids of varying surface tensions $\gamma$, 
from 
38.7\,${\rm Nm^{-1}}$ to 72.2\,${\rm Nm^{-1}}$,
and solids with  
shear modulus $G$ varying from 1.6\,${\rm kPa}$
to 501.9\,${\rm kPa}$. 
In Fig.~\ref{fig:1}a, 
we show a representative experiment   
in which
a 0.5\,mm radius DI water droplet
spreads on a soft substrate with
$G = 1.6\,{\rm kPa}$.
In the initial spreading stage ($t \lessapprox 0.88\,$ms),
the contact radius $r$ increases without interruption 
while the contact line (CL) velocity
gradually reduces from its peak value, 
a behaviour similar to the initial spreading of low viscosity liquids on rigid surfaces 
\cite{biance2004first,winkels2012initial}.
The relatively high CL
velocity ($\sim 1\,$m/s) 
generated by inertia-capillary balance 
leaves the soft surface 
insufficient time to deform and obstruct 
the CL motion \cite{chen2013inertial}. 
In the subsequent stage ($0.88\,\text{ms} \lessapprox t \lessapprox 2.66\,$ms), the liquid spreads through  
several stick-slip cycles 
before it reaches an equilibrium and stops.
Here, sticking occurs when 
either the entire contact line or part of it
is pinned, causing an abrupt drop in CL velocity,
whereas slipping occurs 
when the contact line 
depins
and its velocity resumes.
For instance,  
sticking 
is identified when $r$ appears almost constant (Fig.~\ref{fig:1}a),
although strictly speaking,
the contact line does not completely halt
as it is not simultaneously pinned.
Slipping then occurs when
the contact line progressively liberates itself from the pinned
sites and moves
with a velocity 
significantly higher than that during sticking.

\begin{figure}
\includegraphics[width=0.9\textwidth]{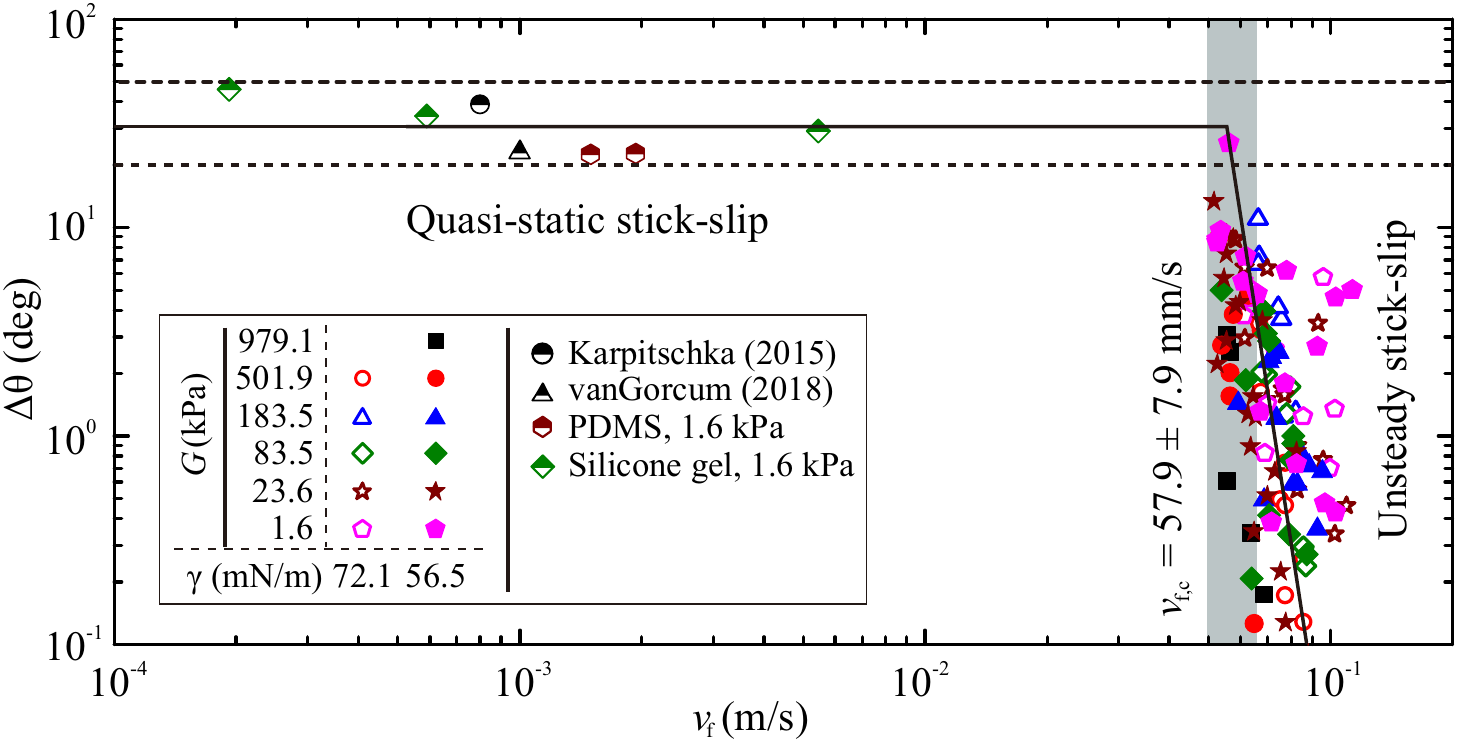}
\centering 
\caption{\small{
Variation of pinning-depinning transition contact angle $\Delta\theta$ on
contact line velocity prior to pinning, $v_{\rm s}$ for 0.5\,mm radius
DI water and 5$\%$ ethanol-water mixture droplets exhibiting 
stick slip motion while 
spreading on soft PDMS 
surfaces with shear modulus
in the range 
$1.6\,{\rm kPa} \leq G \leq 979.1\,{\rm kPa}$. 
Steady state $\Delta\theta$ values
from inflation experiments (see Methods) 
and existing literature  \cite{karpitschka2015droplets,van2018dynamic} are shown for reference.
The dashed lines show the range of values
within which droplets exhibit contact angle hysteresis
on PDMS.}}
\label{fig:2}
\end{figure}

To understand the stick-slip behaviour
in unsteady wetting, 
we explore how it deviates from 
the one observed in quasi-static wetting,
a simplified 
situation
typically associated with low CL velocity.
In a stick-slip cycle of unsteady wetting,
the CL velocity $v$ shifting from negligibly low to high 
is accompanied by 
the corresponding contact angle $\theta$ switching from increasing to decreasing.
As a result, we detect a cyclic relation between 
$\theta$ and $v$ 
(Fig.~\ref{fig:1}b),  
a contrasting behaviour to the 
monotonic increase of 
$\theta$ with $v$ in quasi-static wetting
\cite{van2018dynamic}.
Furthermore, we note that 
CL depinning in quasi-static wetting
is triggered by 
capillary force due to the contact angle increase $\Delta \theta$ developed
during CL sticking
(inset of Fig.~\ref{fig:1}a) \cite{van2018dynamic}.  
In unsteady wetting, however, 
depinning 
is triggered 
by both 
the capillary force 
and the flow inertia 
generated in
the initial fast spreading stage.
Indeed, an assessment of the relation 
between 
$\Delta \theta$ and the flow velocity 
$v_{\rm f}$ representing the flow inertia near the contact line
(Methods)
reveals
a distinction between depinning mechanisms
in quasi-static and unsteady wetting behaviours
(Fig.~\ref{fig:2}).
At small flow velocity ($v_{\rm f} \lessapprox58\,$mm/s), 
i.e., for
quasi-static wetting,
$\Delta \theta$ is approximately a constant comparable to the 
contact angle hysteresis, indicating 
that depinning is only dictated by the contact angle increase 
during sticking, not the 
flow near the contact line. 
This also suggests that the wetting ridge 
is fully developed before depinning \cite{van2018dynamic,gerber2019wetting,park2017self,guan2020state}. 
At high flow velocity ($v_{\rm f} >58\,$mm/s), 
i.e., for
unsteady wetting,
$\Delta \theta$ rapidly drops with increasing $v_{\rm f}$, 
signifying a
formidable role of flow inertia
in causing depinning beside the capillary force.
Moreover,
since the 
contact line depins when $\Delta \theta$ in unsteady wetting
is smaller than that
in quasi-static wetting, 
we infer that depinning occurs in the transient state of wetting ridge formation. 

\begin{figure}
\includegraphics[width=1\textwidth]{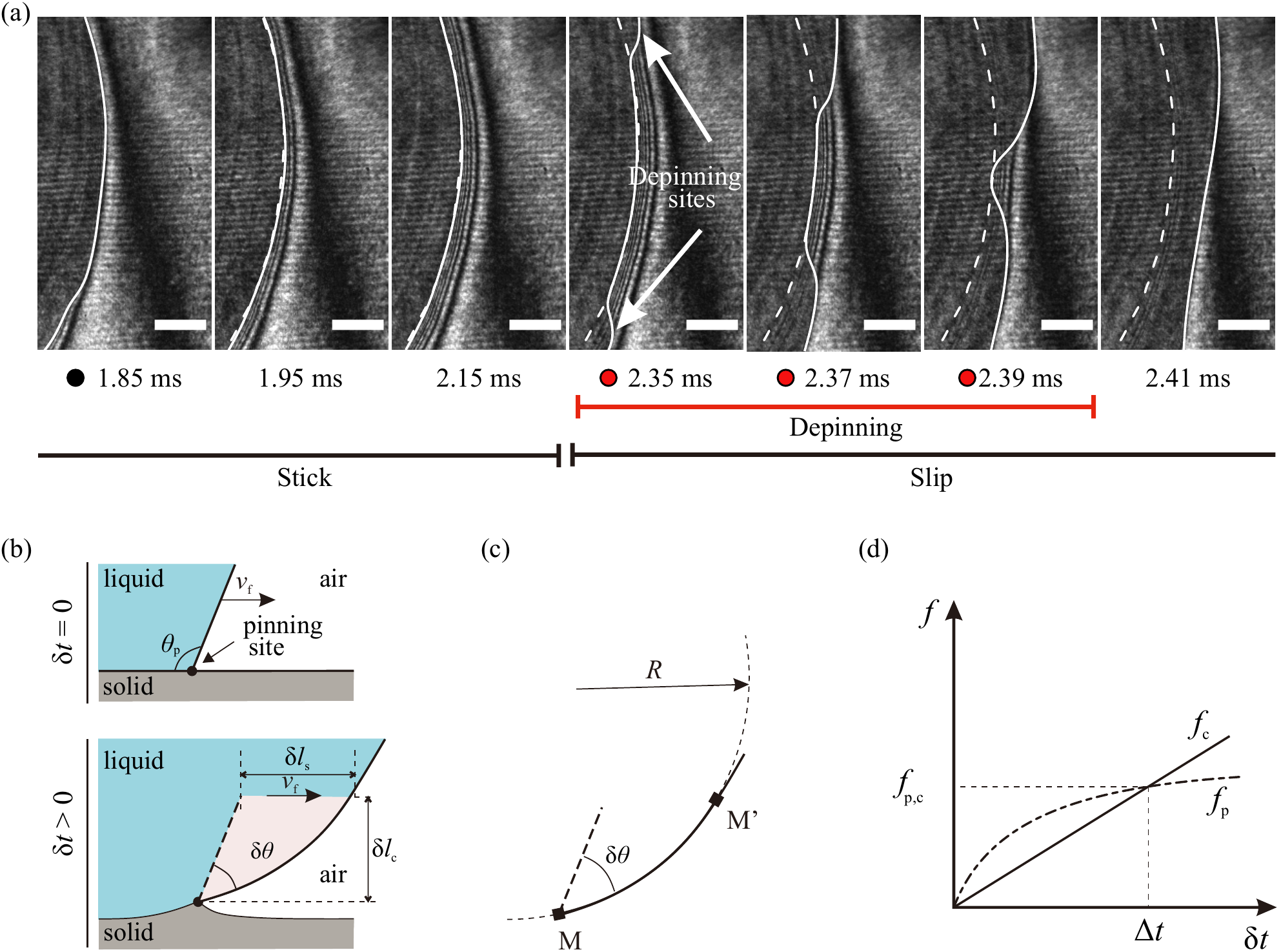}
\centering
\caption{\small{
(a) Snapshots showing the bottom-view interferometric images 
of one typical stick-slip cycles.
The scale bars represent 50$\,{\rm \upmu m}$.
In each snapshot, the solid lines indicate the contact line positions at that moment, while the dashed lines represent the contact line at the pinning moment, i.e., $\delta t = 0$.
(b) Schematics showing dynamical events at contact line during pinning-depinning cycles. 
(c) Force on the liquid-air interface during pinning. 
(d) Illustration showing the change in 
capillary force at the contact line $f_{\rm c}$
and the pinning force 
$f_{\rm p}$ with time $\delta t$.
Depinning happens when the two curves meet, i.e., at $\delta t = \Delta t$
and $f_{\rm p} = f_{\rm p, c}$.}}
\label{fig:3}
\end{figure}

We examine in greater details how depinning occurs when the 
wetting ridge is taking shape 
using bottom-view
interferometric observations,
shown in Fig.~\ref{fig:3}a.
We note that 
a wetting ridge is undetectable before CL pinning, 
but starts growing as soon as CL pinning occurs ($t = 1.85\,{\rm ms}$),
evidenced by the growing number of
interference fringes from $t = 1.85\,{\rm ms}$ to $t = 2.35\,{\rm ms}$
(Fig.~\ref{fig:3}a and Supplementary Fig.~2).
Depinning then is triggered at several weak and
isolated pinned points,
starting from $t = 2.35\,{\rm ms}$
and then completing at $t = 2.41\,{\rm ms}$
when 
the entire contact line 
displaces (Fig.~\ref{fig:3}a).
Nonetheless, 
as the total duration for
depinning is $\approx0.05\,$ms,
much shorter than 
the sticking duration ($\approx0.5\,$ms),
 it is reasonable to assume that 
depinning occurs instantaneously. 
This also implies that depinning is triggered only by  
the liquid flow and contact line conditions during sticking.

As a result, we seek 
to formulate the depinning condition in unsteady wetting
by considering the force balance near the contact line
at a small time $\delta t$ after CL pinning. 
We denote $\theta_{\rm p}$ the contact angle
at pinning, i.e., when $\delta t = 0$. 
As the contact line is fixed but the liquid is still moving due to its inertia, 
the liquid-air interface near the contact line
deforms, causing a contact angle increase 
$\delta \theta$ at $\delta t > 0$ (Fig.~\ref{fig:3}b).
The vertical extent of 
the deformed interface  
is dictated by the capillary velocity $v_{\rm c} \sim \gamma^{1/2} (\rho\,r_0)^{-1/2}$
and is estimated  
as $\delta l_{\rm c} \sim v_{\rm c} \delta t \sin{\theta_{\rm p}}  $.
Here, $\rho$ and $r_0$ are the density and the radius of the droplet, respectively.
The liquid beyond $\delta l_{\rm c}$
continues moving unaffected by CL pinning with the flow velocity $v_{\rm f}$.
As a result, the liquid 
volume $\Omega$ slowed down by 
CL pinning is bounded vertically by $\delta l_{\rm c}$
and horizontally by $\delta l_{\rm s} = v_{\rm f} \delta t$, giving 
$\Omega \sim L \, \delta l_{\rm c} \, \delta l_{\rm s}$, 
where $L$ is the length of the contact line.
We estimate the corresponding deceleration of 
$\Omega$ as $v_{\rm f}/\delta t$. 
This deceleration is caused by 
the capillary force (per unit length)
$f_{\rm c} \sim (\gamma/R) l_{\rm {MM'}}$,
in which the deformed interface has 
the radius of curvature $R$ and length
$l_{\rm {MM'}}\approx (\delta \theta /  \pi) R$ (see Fig.~\ref{fig:3}c).
Thus, the force balance for 
the deceleration of $\Omega$ 
is written as
$L f_{\rm c} \sim \rho \Omega v_{\rm f}/\delta t$, or 
\begin{equation}
\label{eq:deltaTheta5}
\delta \theta
\sim
\pi \sin \theta_{\rm p} {\rm We}\frac{\delta t}{\tau_{\rm c}},
\end{equation}
where 
$\tau_{\rm c} = (\rho r_0^3/\gamma)^{1/2}$ is the inertial-capillary time, 
and 
${\rm We} = v_{\rm f}^2 \rho r_0/\gamma$ is 
a Weber-like number comparing 
the flow inertia 
against capillarity.
The linear relation between $\delta\theta$ and $\delta t$ shown in 
Eq.~\ref{eq:deltaTheta5} is consistent with 
our experimental observations
for different liquids and shear
moduli.

\begin{figure}
\includegraphics[width=1\textwidth]{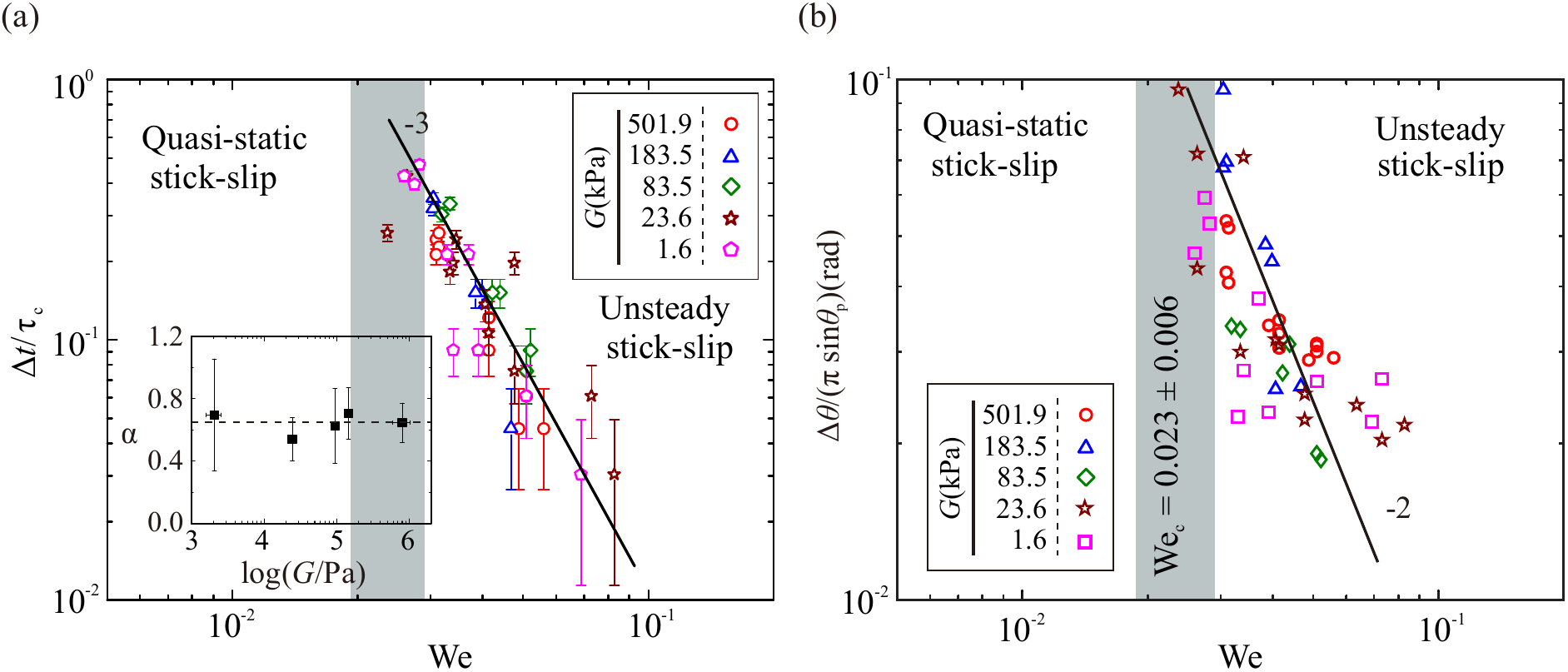}
\centering
\caption{\small{ 
(a) Plot showing $\Delta t/\tau_{\rm c}$ versus ${\rm We}$. 
The solid line shows the scaling law 
$\Delta t/\tau_{\rm c} \sim {\rm We}^{-3}$, equivalently $\alpha = 2/3$.
Inset shows variation of exponent $\alpha$ 
with static shear modulus $G$ 
obtained by fitting the data in (a) to Eq.~\ref{eq:phi}.
$\alpha$ appears to be independent from $G$; 
its average value is 
$\alpha \approx 0.65 \pm 0.07$.
(b) Plot showing $\Delta\theta/(\pi\sin\theta_{\rm p})$ versus ${\rm We}$. 
The solid line shows the
scaling law $\Delta\theta/(\pi\sin\theta_{\rm p}) \sim {\rm We}^{-2}$, 
equivalently $\alpha = 2/3$.
The shaded regions in (a) and (b) 
indicate the critical Weber number ${\rm We_{\rm c}}$ which 
represents the transition between 
quasi-static and unsteady stick slip.}}
\label{fig:4}
\end{figure}

We stress that Eq.~\ref{eq:deltaTheta5}
holds at any moment during sticking, including
the depinning moment, i.e., when $\delta t = \Delta t$
and $\delta \theta = \Delta \theta$.
We recall that for depinning either
on rigid surfaces or on soft surfaces in quasi-static condition, 
the capillary force increase associated with $\Delta \theta$
is required 
to overcome a fixed pinning force 
and is independent of the sticking duration $\Delta t$
\cite{varagnolo2013stick,van2018dynamic}.
On the contrary, 
$\Delta \theta$
consistently increases with $\Delta t$ in our experiment, 
indicating that the CL pinning force
in unsteady soft wetting grows with time. 
Combining with the growth of ridge size during sticking 
(Fig.~\ref{fig:3}a),
we infer that 
the pinning force increases with the pinning time $\delta t$ following 
the growth in ridge size, 
which is finite at $\delta t = 0$. 
Here, the increase in ridge size 
is characterised by the soft solids's 
viscoelastic relaxation timescale
$\tau_{\rm v} = \eta_{\rm s}/G$, where 
$\eta_{\rm s}$
is the dynamic viscosity of the soft solids
\cite{ward2004introduction}.
As a result, we approximate the pinning force (per unit length) 
at $\delta t > 0$
as
$f_{\rm p} \sim k (\delta t/\tau_{\rm v})^{\alpha}$,
where $\alpha$ is a dimensionless 
constant;
$k$ has the dimension of force per unit length and plays a similar role 
to the contact-line spring constant in the 
classical theory of contact angle hysteresis \cite{joanny1984model}.
We note that
the necessary condition to 
depin at 
a finite time 
is $\alpha < 1$ 
following the linear increase of 
$f_{\rm c}$ with $\delta t$
shown in Eq.~\ref{eq:deltaTheta5}.

Depinning, therefore, is possible 
when  
the increase in
capillary force (per unit length)
$\gamma[\cos \theta_{\rm p} - 
\cos{(\theta_{\rm p} + \Delta \theta)}] $
becomes equal 
to 
the 
pinning force (per unit length)
$f_{\rm p}$, or
$\gamma[\cos \theta_{\rm p} - 
\cos{(\theta_{\rm p} + \Delta \theta)}] 
\approx k (\Delta t/\tau_{\rm v})^{\alpha}$.
Combining with Eq.~\ref{eq:deltaTheta5}
at depinning
when $\delta t = \Delta t$
and $\delta \theta = \Delta \theta$, 
the depinning conditions for small $\Delta \theta$ become 
\begin{equation}
\label{eq:depin_time3}
\frac{\Delta t}{\tau_{\rm c}}
\sim 
\Phi\,{\rm We}^{1/(\alpha - 1)},
\,\, {\rm or} \,\,
\frac{\Delta \theta}{\pi \sin \theta_{\rm p}}
\sim
\Phi\, {\rm We}^{\alpha/(\alpha-1)},
\end{equation}
where the dimensionless parameter $\Phi$ is 
\begin{equation}
\label{eq:phi}
\Phi =
\left(
\frac{\pi \gamma \sin^2 \theta_{\rm p}}{k} \right)^{1/(\alpha - 1)}
\left(\frac{\tau_{\rm v}}{\tau_{\rm c}}\right)^{\alpha/(\alpha - 1)}.
\end{equation}
We highlight that 
Eq.~\ref{eq:depin_time3}
explicitly predicts 
both the maximum duration $\Delta t$ 
and 
the maximum contact angle increase $\Delta \theta$ 
from pinning to depinning. 
We experimentally test these predictions 
by showing 
$\Delta t/\tau_{\rm c}$ 
and $\Delta \theta/(\pi \sin \theta_{\rm p})$
versus
${\rm We}$ 
in Fig.~\ref{fig:4}a 
and Fig.~\ref{fig:4}b, respectively.
The experimental data, comprising of 
all substrates with 
$G$ varying from 1.6\,kPa to 501.9\,kPa, 
unequivocally agree with both
power laws derived in Eq.~\ref{eq:depin_time3}
for $\alpha = 0.65 \pm 0.07$. 
The inset of Fig.~\ref{fig:4}a
shows that $\alpha$ 
fluctuates around 0.65 (dashed line in the inset)
for $G$ varying almost three orders of magnitude,
indicating that $\alpha$ 
can be practically considered 
constant in the explored range of shear modulus.
The shaded areas in Fig.~\ref{fig:4} 
indicate the critical Weber number ${\rm We_{\rm c}} = 0.023 \pm 0.006$ 
beyond which unsteady stick-slip happens 
(Methods).
This excellent agreement 
thus confirms our theoretical model on the depinning condition for 
unsteady wetting on soft surfaces. 

\section*{Discussions}

\subsection{Dynamic defects and soft contact line elasticity.}
The scaling law describing the pinning force on soft solids
at the contact line (CL),
$f_{\rm p} \sim k (\delta t/\tau_{\rm v})^\alpha$ with
$\alpha = 0.65 \pm 0.07 \approx 2/3$,
is consistent with the classical theory on
contact-line pinning \cite{joanny1984model}
considering growing deformation on soft solids 
as dynamic surface heterogeneities.

For wetting on rigid surfaces, 
the widely accepted theory 
for contact-line pinning \cite{joanny1984model,bonn2009wetting,vo2019critical} 
attributes surface defects of a fixed characteristic size $\zeta$
to causing CL pinning and subsequent
deformation of the liquid-air interface. 
The maximum pinning force (per unit length) before depinning 
is calculated as
$f_{\rm p} \sim k_{\rm r} \varepsilon_{\rm m}$, 
where the parameter $k_{\rm r} \sim \pi \gamma \sin^2 \theta_{\rm p}/\ln (L'/\zeta)$
acts as a spring constant for elastic 
deformation of the liquid-air interface near the contact line.
Here, $L'$ is the
upper-bound length scale of
the CL region and is conveniently estimated using droplet size
in the case of wetting by liquid droplets;
the normalised amplitude $\varepsilon_{\rm m}$ at maximum deformation 
is 
estimated as 
$\varepsilon_{\rm m} \sim (\zeta/L')^{\beta}$, where
the exponent $\beta$ is either $1/2$
for $\varepsilon_{\rm m} < \zeta/L'$, 
or $2/3$
for $\varepsilon_{\rm m} > \zeta/L'$.

For unsteady wetting on soft surfaces,
the defect size is not constant
due to size variation of wetting ridges. 
The growth of the wetting ridge near the depinning moment
can be expressed
using Kelvin-Voigt model \cite{ward2004introduction}
as $\zeta = l [1 - \exp (-\delta t/\tau_{\rm v})]$ for $0<\delta t\lessapprox \Delta t$.
Since the forcing duration $\delta t$ during CL pinning
is a few orders of magnitude
smaller than the viscoelastic relaxation time, i.e., $\delta t \ll \tau_{\rm v}$,
we simplify the expression for $\zeta$ using Taylor expansion as
$\zeta \approx l(\delta t/\tau_{\rm v})$.
Using this expression for the defect size
leads to a modified pinning force (per unit length) near the depinning moment
for soft wetting:
\begin{equation}
\label{eq:pinning_force2}
f_{\rm p} 
\sim k_{\rm r} 
\left(\frac{l}{L'} \right)^{\beta} \left(\frac{\delta t}{\tau_{\rm v}}\right)^{\beta}.
\end{equation}
Here, we note that $k_r$ 
is reasonably considered as a time-independent parameter for $0<\delta t \lessapprox \Delta t$.
We highlight that 
by setting $\beta = \alpha$ and 
$k =k_{\rm r} (l/L')^{\alpha}$,
Eq.~\ref{eq:pinning_force2} yields 
the expression for the pinning force (per unit length) used
in our calculation of the depinning condition for soft wetting
\begin{equation}
f_{\rm p} \sim k\left(\frac{\delta t}{\tau_{\rm v}}\right )^{\alpha}.
\end{equation}

The time-independent constant $k$, herein referred to as  
the CL \emph{soft spring constant},
is related to the material properties of the involved phases.
A direct consequence of the relation between 
$k$ and its rigid counter-part $k_{\rm r}$
(noting that typically $l/L' <1$)
is that soft surfaces most likely have lower spring constant
-- or higher contact angle hysteresis --
than that on rigid surfaces.
Moreover,
we can safely assume 
that the maximum deformation of the liquid-air interface
$\varepsilon_{\rm m}L'$ 
is much larger 
than the defect size $\zeta$ of soft solid.
As a result, we obtain $\varepsilon_{\rm m} > \zeta/L'$, or $\varepsilon_{\rm m} > \zeta/L'$,
which leads to $\beta = 2/3$ \cite{joanny1984model}, 
consistent with our experimentally determined 
exponent $\alpha = 0.65 \pm 0.07$.


\section*{Conclusion}
From our experiments of unsteady wetting on 
soft solids and the accompanying theoretical analysis, 
we find that unsteady wetting on soft solids
exhibits
intermittent spreading behaviour,
i.e., the contact line (CL) alternates between 
sticking and slipping,
when the
liquid's spreading rate becomes comparable to 
the soft solid's 
viscoelastic reacting rate.
During intermittent spreading, the  
contact line slips or sticks
following the winning factor between 
the liquid's inertia and the pinning force at the contact line. 
Setting the transient inertia of the liquid during CL pinning 
equal to the opposing pinning force that 
grows at the same rate as the 
solid's viscoelastic characteristic timescale, 
we obtain the predictive 
condition for CL depinning,
including 
the maximum 
time and the maximum contact angle increase 
allowed during CL pinning. 
This is experimentally verified and further supported 
by a modification of 
the classical theory for CL pinning on  
undeformable surfaces, 
treating the 
growing wetting ridge as dynamic surface heterogeneities.




\section*{Methods}
\subsection{Fabrication and characterization of soft surfaces.}
Soft surfaces were fabricated using PDMS 
(Polydimethylsiloxane, Sylgard 184, Dow) 
with the polymer-crosslinker weight ratios: 
9:1, 10:1, 20:1, 30:1, 40:1, and 60:1
First, a PDMS mixture was prepared, 
stirred using a magnetic stirrer for 30 minutes, 
and degassed in a vacuum chamber 
for 60 minutes. 
The degassed mixture was coated 
on freshly cleaned microscope glass slides (Witeg)
using a spin coater (POLOS 200, SPS)
at a spin rate 2000 rounds-per-minute (RPM)
for one minute to obtain a coating thickness of 75\,$\mu$m. 
The PDMS substrate was finally cured in an oven at $70^{\circ}$ for 8 hours. 
The elastic properties of fabricated soft substrates 
were characterised using
a rheometer (Discovery HR-20, TA Instruments). 
The surface roughness was measured by 
an atomic force microscope (AFM, Bruker) 

\subsection{Unsteady wetting experiment and imaging.}
We use DI water and ethanol-water mixtures as our working liquids. 
The schematic of the experimental setup is provided in
the Supplementary Fig.~8.
Droplets were generated at the tip of a stainless steel needle 
by injecting the liquid at a small rate ($1{\rm {\mu}L\,min^{-1}}$). 
When the radius of a droplet reached $r_{0} = 0.5$\,mm,
its lower surface gently touched a soft surface and the liquid started
spreading. 
The spreading dynamics was recorded from the side
using a high-speed camera (SA-X2, Photron)
running at 50,000 frames per second (FPS),
with back illumination provided 
by a diffused metal halide light source (LS-M180, Sumita). 
The side-view camera was equipped with a
20X 
long-working-distance objective (M-Plan Apo 20X, Optem Engineering),
providing a resolution of 
$1\,\mu{\rm m}/\text{pixel}$.
Bottom-view interferometric recordings of the wetting ridge
were obtained by a high-speed camera
(SA-X2, Photron)
imaging at 50,000\,FPS.
The camera was coupled with a 
20X objective, providing 
a resolution of 0.9\,$\mu$m/pixel. 
Coaxial illumination
was provided by a
green laser of wavelength $\lambda = 532$\,nm
and a beam-splitter. 
The interferometric patterns indicating surface deformation, i.e., the wetting ridge
were formed by reflected light from the the glass-PDMS interface
and the PDMS-air interface on the dry side of the contact line. 
From the recorded interferometric patterns, 
deformation caused by the wetting ridge 
outside of the liquid droplet
is reconstructed.

\subsection{Determination of contact radius and contact angle.}
In side-view recordings of spreading droplets, the two 
points representing the CL positions
on the side of a droplet 
are those where 
the air-liquid interface appears to intersect with the solid surface. 
These two points, herein referred to as side-view contact-line points
(or side CL points), 
are used to obtain the contact radius $r$
and its time dependence.

We note that  
the contact line 
in realistic wetting situations
often does not stick nor slip
synchronously. 
Tracking the side CL points
is essentially tracking the faster-moving 
points of 
the contact line within the observing volume of the lens
used to obtain the side-view images.
This is consistent with our experimental observations
showing that even during sticking, 
the velocity $v_{\rm p}$
of the side CL points never falls to zero.
This also implies a small fluctuation 
of the contact angle $\theta$
along the contact line at any given time;
and a non-zero deviation 
between $\theta$
and the contact angle $\theta_{\rm m}$
measured 
from side-view recordings.
The deviation between $\theta$
and $\theta_{\rm m}$
is negligible during slipping, 
but it is
significant during sticking 
where the side CL point appears 
moving slowly with velocity $v_{\rm p}$
while its neighboring points are  
pinned and not observable
from the side view
(Fig.~\ref{fig:3}a).
It is possible to relate
$\theta$ to $\theta_{\rm m}$ during sticking
as follows: 
\begin{equation}
\label{eq:theta_correction}
\tan \theta = \frac{v_{\rm c} }{v_{\rm c}  - v_{\rm p} \tan \theta_{\rm m}} \tan \theta_{\rm m}.
\end{equation}

\subsection{Flow velocity near contact line.}
We note a distinction between 
the contact line velocity $v$
and the fluid flow velocity $v_{\rm f}$ near the contact line, i.e.,  
$v$ is the instantaneous CL velocity resulted 
from a complex interaction between flow inertia, capillary force, and 
the surface's elasticity at the contact line, 
whereas $v_{\rm f}$ is
only governed by the flow inertia accumulated in the 
fast initial spreading stage.
The dependence of $v_{\rm f}$ on the initial flow inertia
is consistent with the observation that 
the spreading
radius strictly obeys the power law $r \sim t^{b}$
in the initial fast spreading stage and 
subsequently fluctuates around this law 
in the stick-slip stage due to the alternative stick-slip cycles
\cite{chen2011short}.
As a result, 
we define the flow velocity $v_{\rm f} = C_{0}t^{b-1}$
as the one characterising
the flow inertia accumulated in the 
fast initial spreading stage. 
Here, the exponent $b$, varying from 0.25 to 0.3, 
results from 
fitting the power law to the data \cite{chen2011short};
the constant $C_{0}$, varying from $1.4 \times 10^{-3}$ to $2.5 \times 10^{-3}$,
is determined 
for each spreading experiment using the
least-square fitting method.

\subsection{Critical Weber number for unsteady stick-slip behaviour.}
There exists a critical Weber number, denoted as ${\rm We_c}$, 
separating 
the quasi-static stick-slip and 
unsteady stick-slip flow behaviours. 
The critical Weber number 
is estimated when the ridge's growing duration during pinning, $\Delta t$ 
is comparable to the visco-elastic relaxation timescale $\tau_{\rm v}$ of the solid.
As a result, by setting $\Delta t = \tau_{\rm v}$ in Eq.~\ref{eq:depin_time3}, 
we obtain
\begin{equation}
\label{eq:vL}
{\rm We_c} \sim \left( \frac{\tau_{\rm c}}{\tau_{\rm v}} \Phi \right)^{1-\alpha}
=
\frac{k}{\pi \gamma \sin^2 \theta_{\rm p}} \frac{\tau_{\rm c}}{\tau_{\rm v}}.
\end{equation}
Subsequently, using the expression for soft spring constant
$k \sim \pi \gamma \sin^2 \theta_{\rm p}(l/L')^{\alpha} [\ln (L'/\zeta)]^{-1}$,
we obtain an explicit expression 
of the critical Weber number:
\begin{equation}
\label{eq:vL1}
{\rm We_c} \sim 
[\ln (L'/\zeta)]^{-1}
\left(\frac{l}{L'} \right)^{\alpha}
\frac{\tau_{\rm c}}{\tau_{\rm v}}.
\end{equation}
Our experimental data shows that 
${\rm We_c} \approx 0.0234 \pm 0.006$
for the explored range of shear modulus.
This translates to the critical flow velocity 
$v_{\rm f,c}\approx 57.9 \pm 8\,$mm/s.
The shaded areas in Fig.~\ref{fig:2} 
and Fig.~\ref{fig:4} respectively 
represent the critical velocity and the critical Weber number,
consistent with the experimentally observed 
transitions between quasi-static and unsteady stick-slip regimes.


\section*{References}
\bibliography{stickSlip_SM_QV_ML_TT}


\begin{addendum}
\item This study is supported by
Nanyang Technological
University, 
the Republic of Singapore's Ministry of Education 
(MOE, Grant No. MOE2018-T2-2-113), 
and the RIE2020 Industry Alignment Fund – Industry Collaboration Projects (IAF-ICP) Funding Initiative, as well as cash and in-kind contribution from the industry partner, HP Inc. 
S.M. is supported by NTU Research Scholarship.
M.L. is supported by Nanyang President’s Graduate Scholarship.
\item[Author contributions statement]
T.T. conceived the study. 
S.M. and Q.V. performed the experiment and analysed the data with assistance from M.L. S.M., Q.V., and T.T. discussed the results, T.T. wrote the manuscript with inputs from S.M., Q.V., and M.L; T.T. supervised the research.

\item[Competing Interests]
The authors declare that they have no
competing financial interests.
\item[Correspondence]
Correspondence and requests for materials
should be addressed to
Tuan Tran~(email: ttran@ntu.edu.sg),
\end{addendum}


\end{document}